\documentstyle[sprocl,bugfix]{article}

\input{psfig}
\begin{document}

\title{A Haplousterotic Model for the Nucleon Wave Function}

\author{R. Eckardt, J. Hansper, M.F. Gari}

\address{Institut f\"ur Theoretische Physik, Agr.\ Mittelenergiephysik,\\
         Ruhr-Universit\"at Bochum, D-44780 Bochum, Germany\\
         E-Mail: QCD-MEP@MEP.Ruhr-Uni-Bochum.de}


\def\Journal#1#2#3#4{{#1} {\bf #2}, #3 (#4)}

\def\NCA{\em Nuovo Cimento}
\def\NIM{\em Nucl. Instrum. Methods}
\def\NIMA{{\em Nucl. Instrum. Methods} A}
\def\NPB{{\em Nucl. Phys.} B}
\def\PLB{{\em Phys. Lett.}  B}
\def\PRL{\em Phys. Rev. Lett.}
\def\PRD{{\em Phys. Rev.} D}
\def\ZPC{{\em Z. Phys.} C}
\def\st{\scriptstyle}
\def\sst{\scriptscriptstyle}
\def\mco{\multicolumn}
\def\epp{\epsilon^{\prime}}
\def\vep{\varepsilon}
\def\ra{\rightarrow}
\def\ppg{\pi^+\pi^-\gamma}
\def\vp{{\bf p}}
\def\ko{K^0}
\def\kb{\bar{K^0}}
\def\al{\alpha}
\def\ab{\bar{\alpha}}
\def\be{\begin{equation}}
\def\ee{\end{equation}}
\def\bea{\begin{eqnarray}}
\def\eea{\end{eqnarray}}
\def\CPbar{\hbox{{\rm CP}\hskip-1.80em{/}}}


\maketitle\abstracts{%
We review the consequences of the sensitivity of the relation between the
 {\em moments\/} of a model for the nucleon quark distribution amplitude
 and the {\em coefficients\/} of its polynomial expansion.
Criteria for a simpler approach to constructing a model for the quark
 distribution amplitude are formulated. 
We describe how such a simpler (or {\em haplousterotic\/}) model for the quark
 distribution amplitude of the nucleon is obtained from the QCD sum-rule
 moments of COZ.}

\section{Introduction}

\begin{figure}
\def\EpsfFourBox#1,#2{%
    \hbox to 1.17 in{%
        \hglue 0.595 in \dimen0 = #1 bp \dimen0=0.333\dimen0
        \hglue \dimen0
        \hbox to 0pt{\hss\psfig{file=#2,width=1.8in,silent=}\hss
        }\hss
    }\ignorespaces
}
\def\TextFourBox#1,#2{%
    \hbox to 1.17 in{%
        \hglue 0.595 in \dimen0 = #1 mm \dimen0=0.333\dimen0
        \hglue \dimen0 \hbox to 0pt{\hss #2 \hss}\hss
    }\ignorespaces  
}
\def\phiN#1{{\def\arg{{#1}}\ifmmode{\phi_N^\arg}\else $\phi_N^\arg$\fi}}
\def\phias{\phi_{\rm as}}
\vbox{
\vglue  5.4 mm
\hbox{%
   \hskip 2.45 cm
   \TextFourBox   0,{$\phias$}
   \hskip 1.4 cm
   \TextFourBox -18,{$\phiN{\rm COZ}$}
   \hskip 2.45 cm}
\vskip -9 mm
\hbox{%
   \hskip 2.45 cm
   \EpsfFourBox 0,{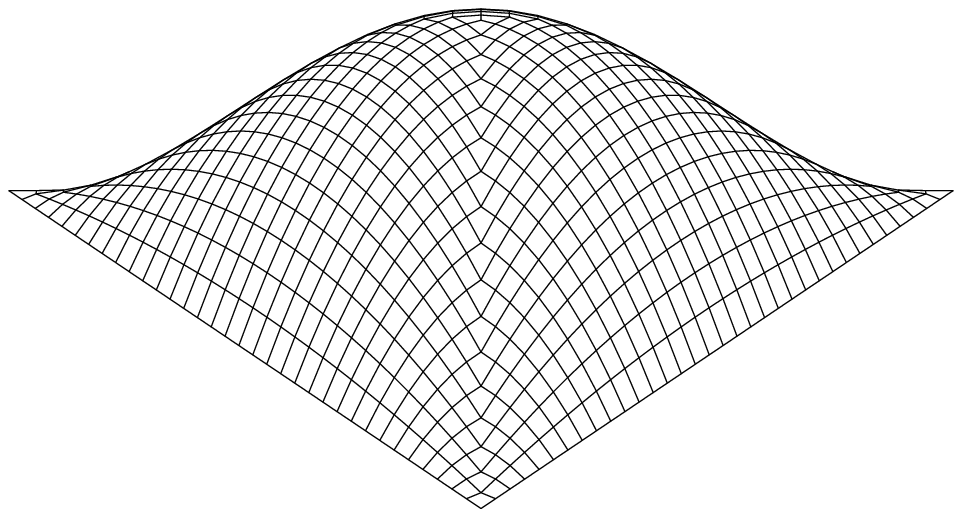}
   \hskip 1.4 cm
   \EpsfFourBox 0,{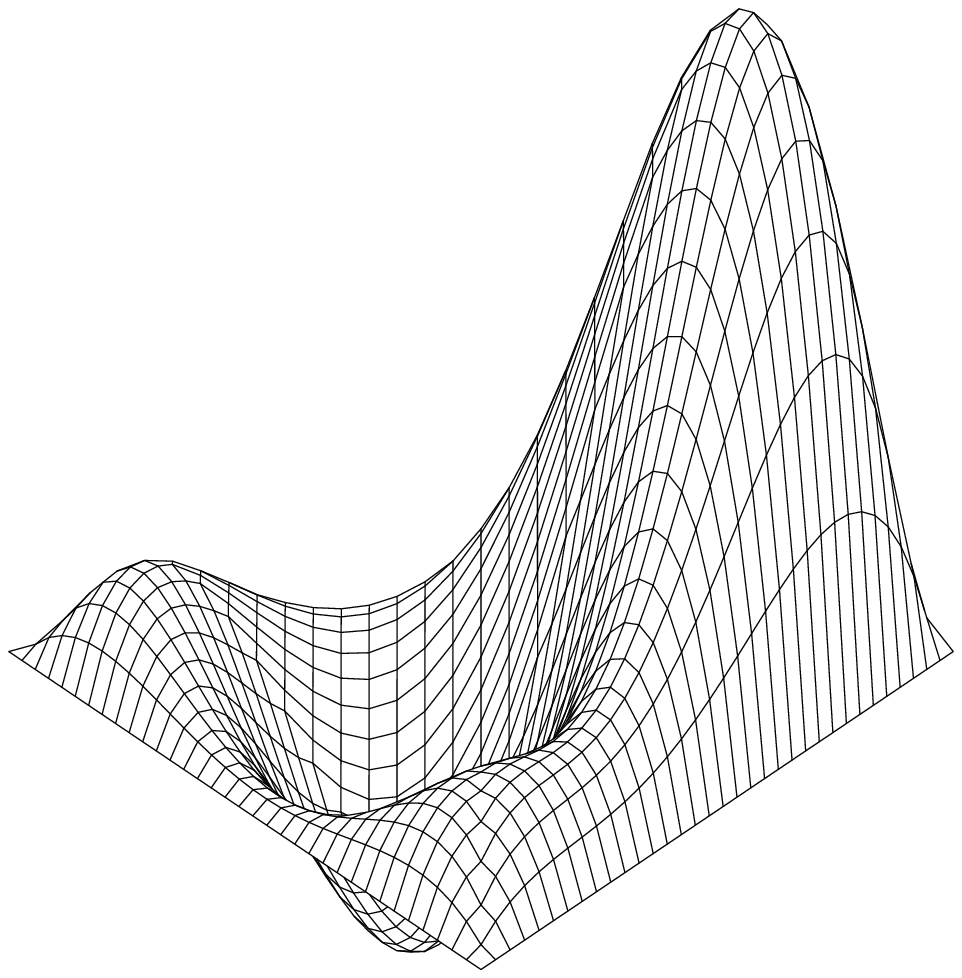}
   \hskip 2.45 cm
}
\vglue -3.6 mm
\hbox{%
   \TextFourBox -16,{$\phiN{\rm CZ}$}
   \hskip 1.33 cm
   \TextFourBox -13,{$\phiN{\rm KS}$}
   \hskip 1.69 cm
   \TextFourBox -23,{$\phiN{\rm GS}$}}
\vskip -6.3 mm
\hbox{%
   \EpsfFourBox 0,{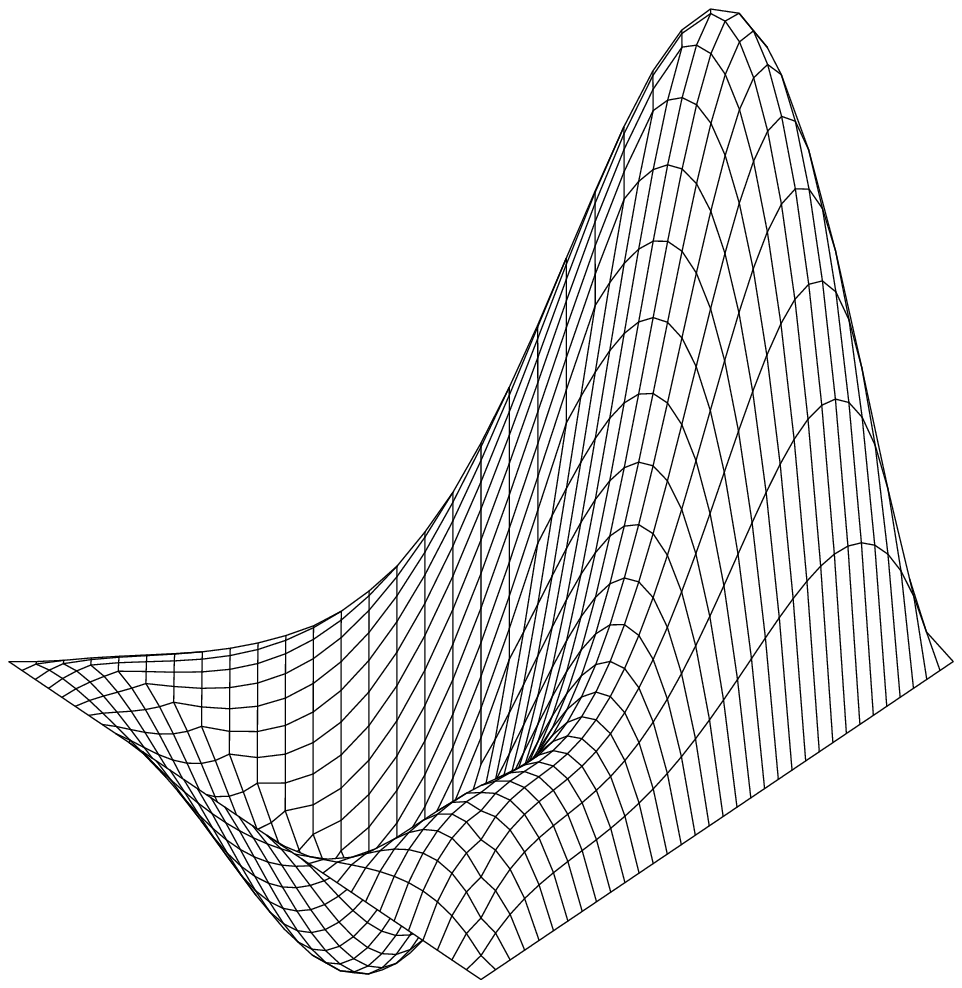}
   \hskip 1.33 cm
   \EpsfFourBox 0,{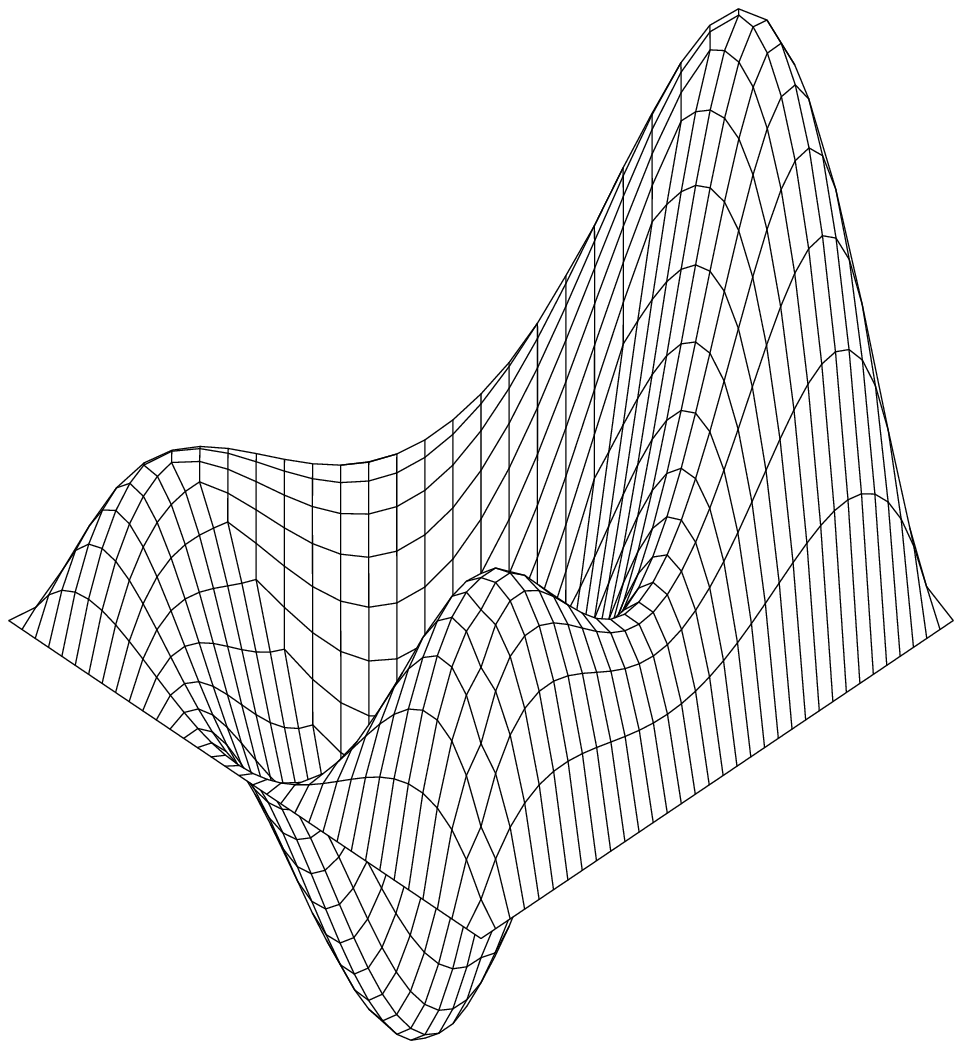}
   \hskip 1.69 cm
   \EpsfFourBox 0,{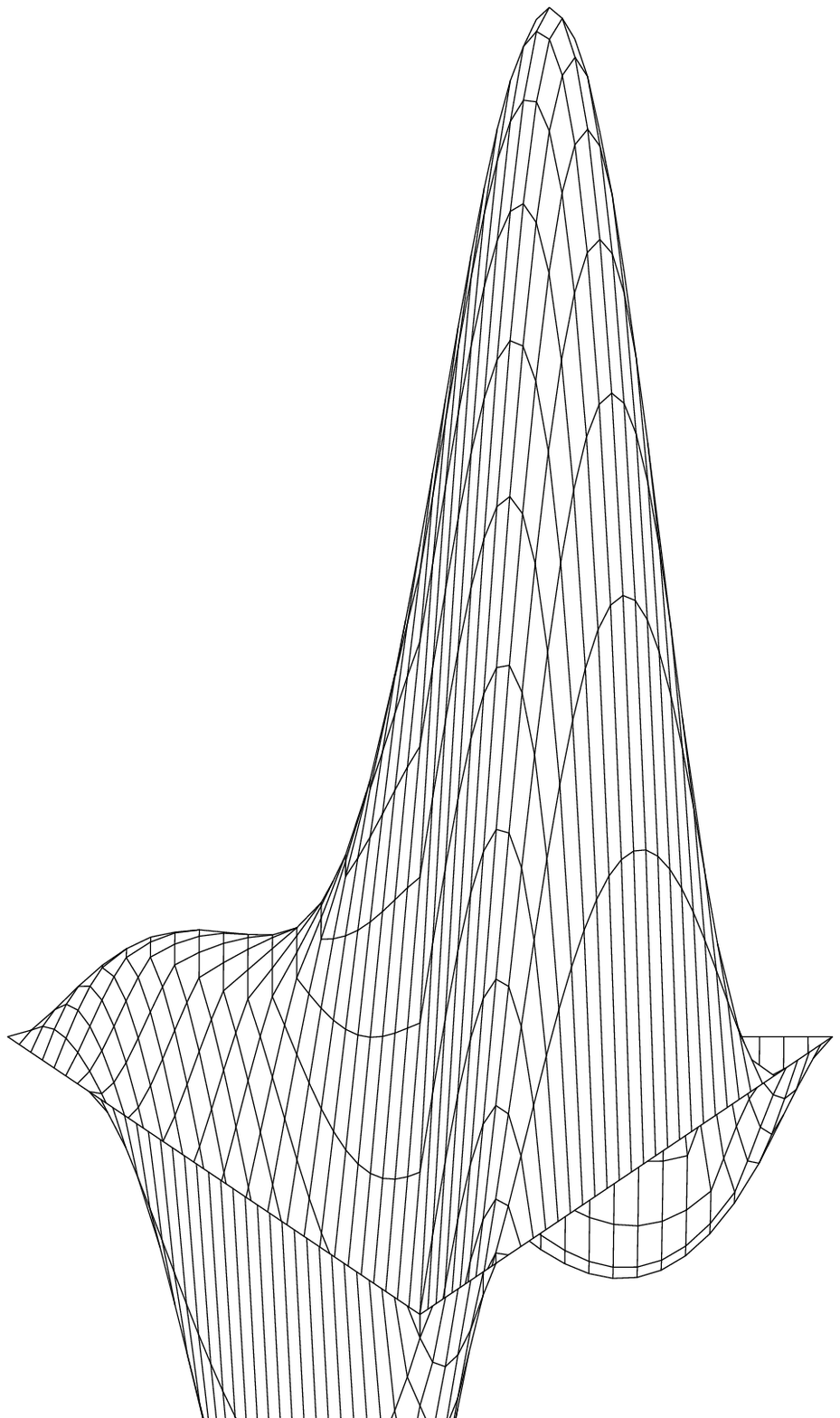}
}
\vglue -3.6 mm
\hbox{%
   \hskip 6.20 cm
   \TextFourBox -28,{$\phiN{\rm het}$}
   \hskip 2.85 cm}
\vskip -4.5 mm
\hbox{%
   \hskip 6.20 cm
   \EpsfFourBox 0,{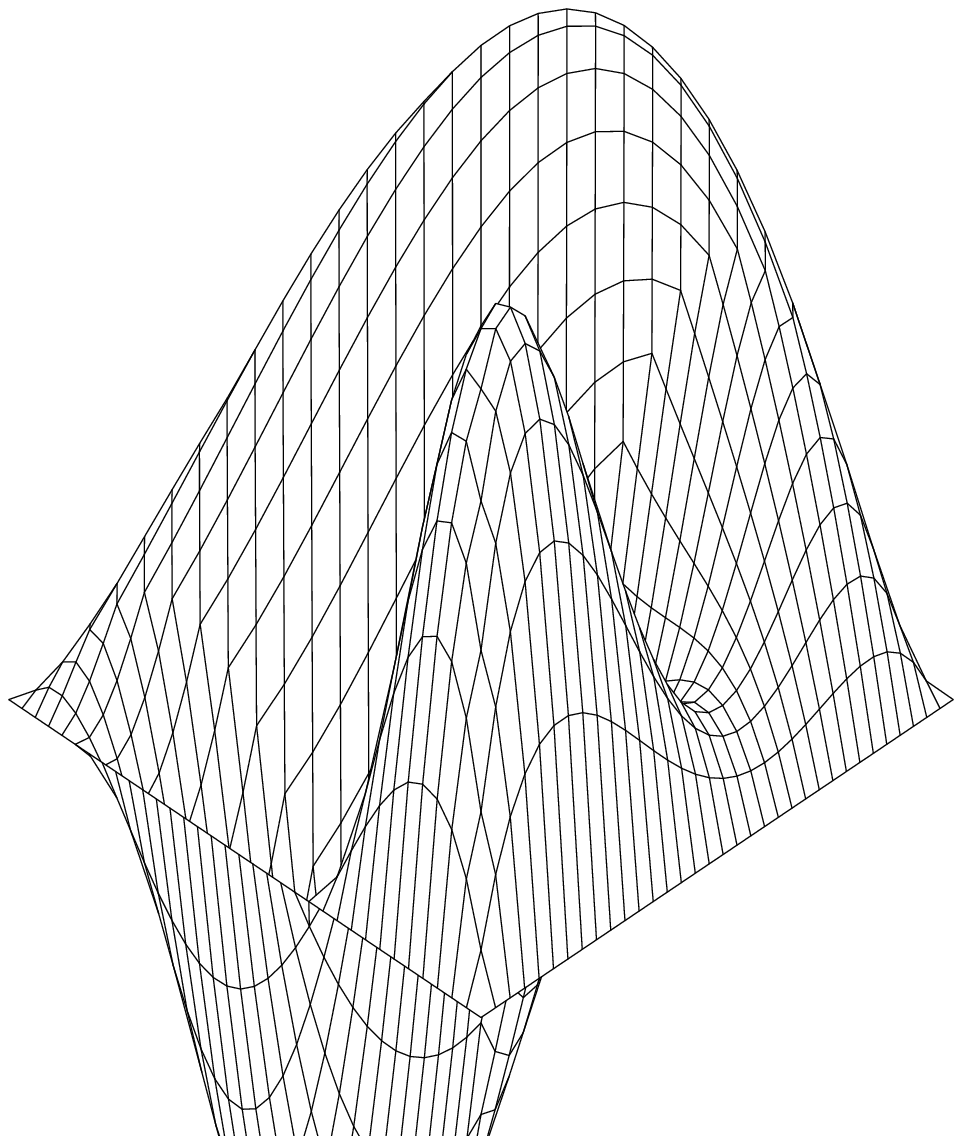}
   \hskip 2.85 cm
}
\vglue -22.5 mm
\hbox{%
   \TextFourBox  18,{$\phiN{\rm III}$}
   \hskip 6.23 cm}
\vskip 0.9 mm
\hbox{%
   \EpsfFourBox 0,{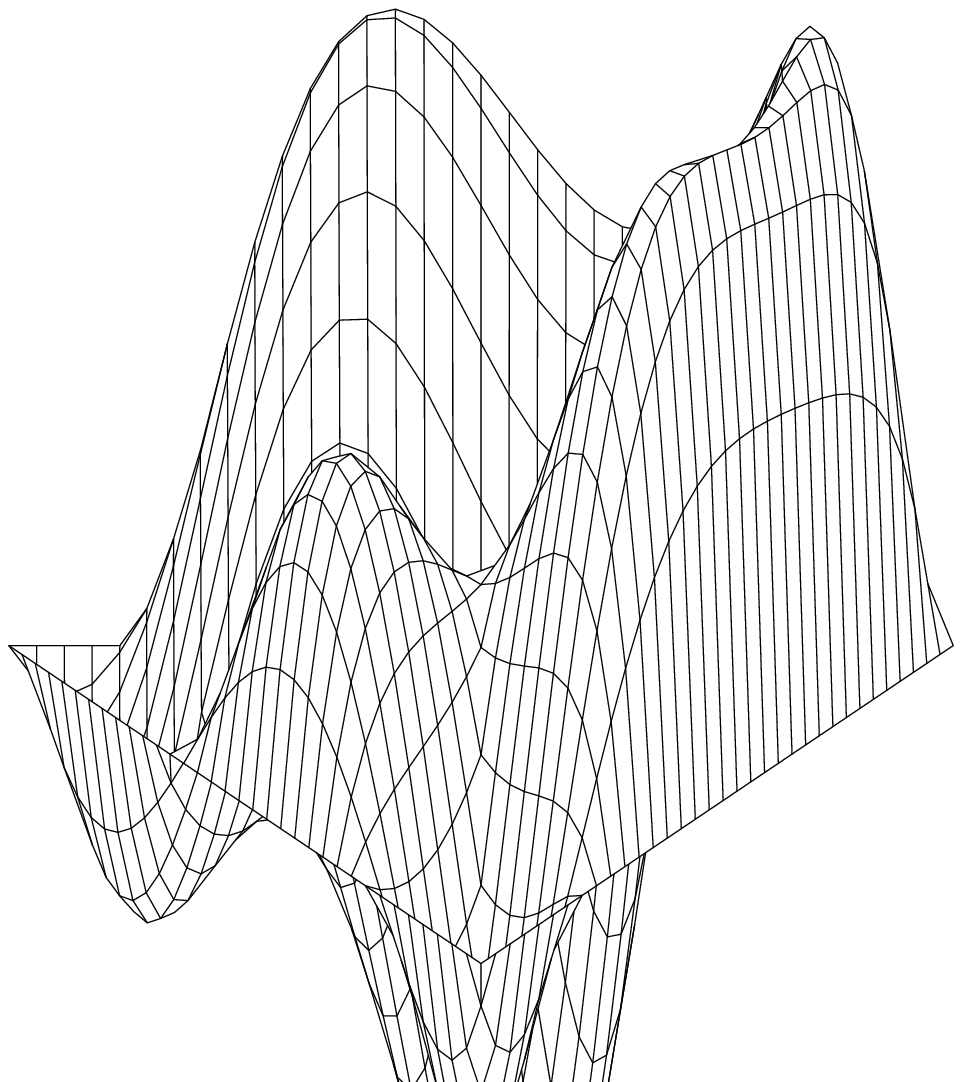}
   \hskip 6.23 cm
}
\vglue -25 mm
\hbox{%
   \hskip 3.11 cm
   \TextFourBox   7,{$\phiN{\rm IVc}$}
   \hskip 2.91 cm
   \TextFourBox   6,{$\phiN{(1)}$}}
\vskip -2.7 mm
\hbox{%
   \hskip 3.11 cm
   \EpsfFourBox 0,{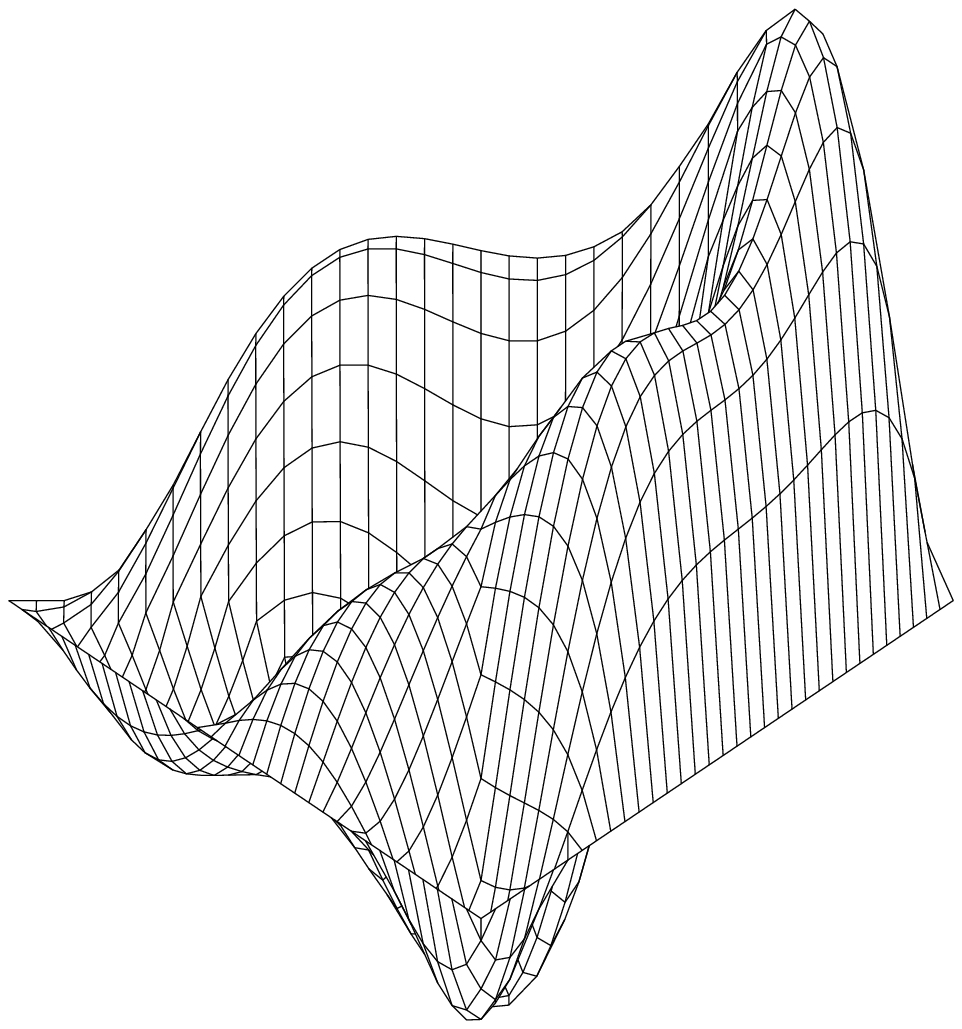}
   \hskip 2.91 cm
   \EpsfFourBox 0,{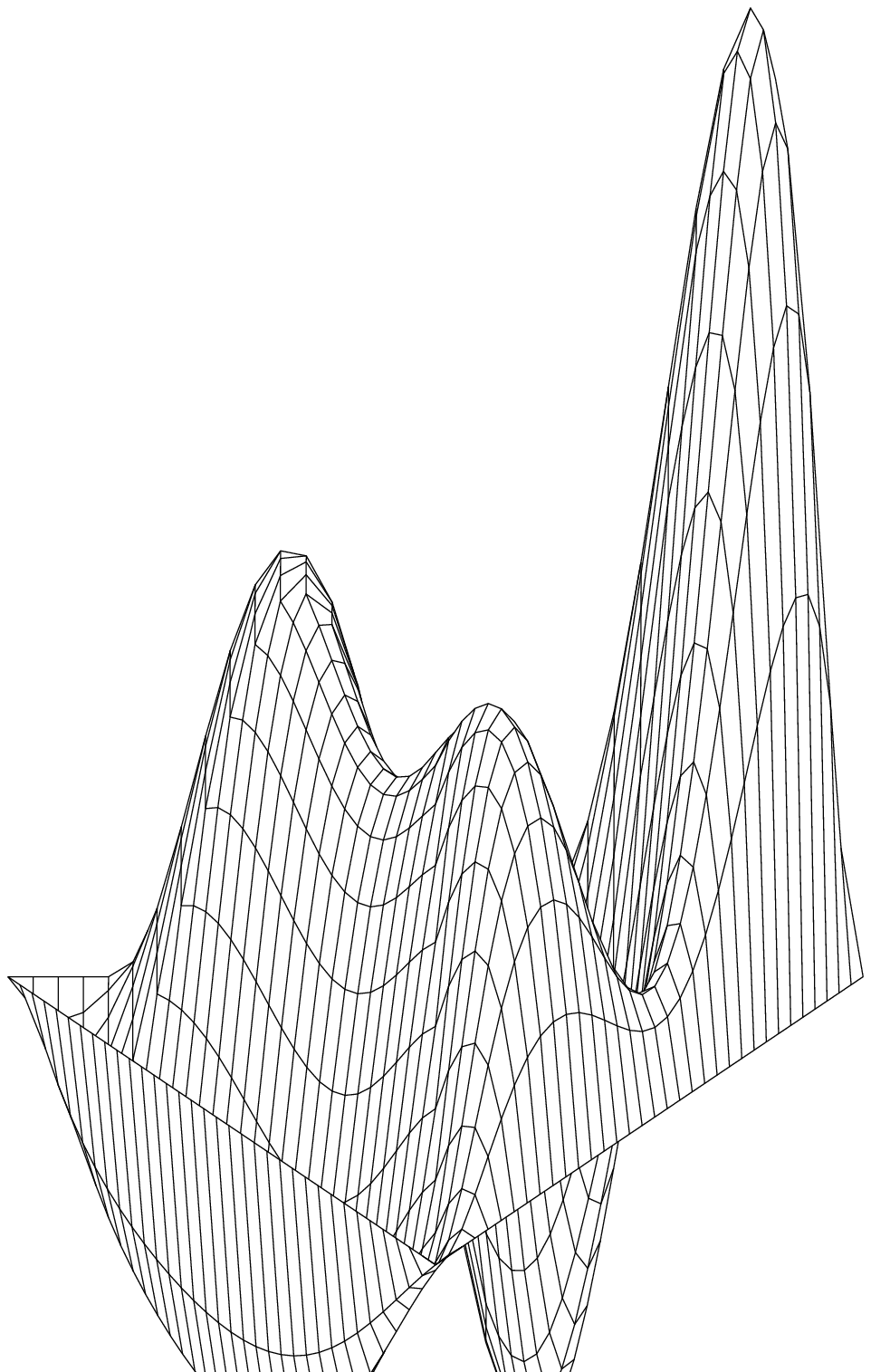}
}
\vskip 5.4 mm
}%
\vskip -3 mm
\caption{%
Some selected models for the nucleon wave function.
The model amplitudes are polynomial approximations of second and
 third degree from Refs.\ \protect\citelow{COZ89} (COZ),
 \protect\citelow{CZhs84} (CZ),
 \protect\citelow{KiSa87} (KS), \protect\citelow{GaSt87} (GS),
 \protect\citelow{StBe93} (het), \protect\citelow{Schae89} (III, IVc),
 and \protect\citelow{HEG92} (1).
$\phi_{\rm as}=120\,x_1 x_2 x_3$ denotes the asymptotic form.
\label{FIGModelle}}
\end{figure}

Although the knowledge of the nucleon wave function in terms of its
 fundamental
 quark and gluon degrees of freedom is of outstanding theoretical
 interest because of its {\em process-independence},
 there are many drastically different model amplitudes for the nucleon
 available as shown in Fig.~\ref{FIGModelle}.
All the functions shown are intended to describe the distribution of the
 longitudinal momentum fractions of the quarks inside the nucleon.

\begin{figure}
\centerline{\psfig{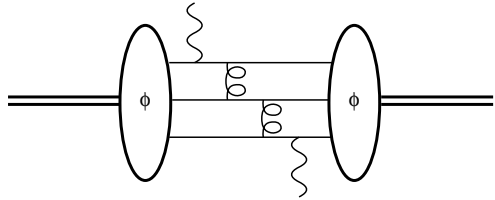}}
\vskip -2 mm
\caption{%
A typical lowest-order contribution to virtual compton scattering in the
 hard-scattering picture of pQCD.
While the hard gluon exchanges can be calculated perturbatively, the nucleon
 QDA~$\phi$ that contains all the non-perturbative information is usually
 obtained from QCD sum-rules.
\label{VCS-contr}}
\end{figure}

In calculations using perturbative Quantum Chromodynamics (pQCD), not
 the full wave function $\psi_N$ itself is used, but the quark distribution
 amplitude (QDA) $\phi_N$, which is defined from the nucleon wave function by
\begin{equation}
  \phi_N(x,\mu^2):=
  \int_{\vec{k_\perp}<\mu^2} [d^2{k_\perp}_i]\; \psi_N(x_i,\vec{k_\perp}_i)
  \quad .\label{QDA}
\end{equation}
From Eq.~\ref{QDA} is clear that $\phi_N$ is also process-independent and can
 thus be used in calculations of various observables such as formfactors, decay
 widths or amplitudes in virtual Compton scattering\cite{FaZh90} (VCS).
In the hard-scattering aproach\cite{LB80} of pQCD the contributions to such
 processes are assumed to factorize into a process-specific hard-scattering
 amplitude $T_H$ between the distribution amplitudes of in- and outgoing
 states.
For example, a typical contribution to $T_H$ in VCS looks like the diagram in
 Fig.~\ref{VCS-contr}.

Whereas the hard-scattering amplitude can be calculated perturbatively
 by exchanging hard gluons, the QDA $\phi_N$ is a purely non-perturbative
 object and must be obtained from elsewhere.

Usually, the ansatz for $\phi_N$ was an expansion into a set of orthogonal
 polynomials (Appell polynomials $A_i(x_1,x_2,x_3)$ for the
 nucleon\cite{LB80,Appell}).
The remaining task was to find the corresponding expansion coefficients $c_i$.
From the definition of the moments of~$\phi_N$
\begin{eqnarray}
\langle i j k\rangle
  :=\langle x_1^i x_2^j x_3^k\rangle
 &:=&\int\thinspace[dx]~ x_1^i\; x_2^j\; x_3^k\cdot
       \phi_N(x_1,x_2,x_3)\hfill                      \label{NUKMOM}\\
 &=& \sum_{i=0}^\infty c_i \cdot\int\thinspace[dx]~
    x_1^i\; x_2^j\; x_3^k\cdot A_i(x_1,x_2,x_3)~, \nonumber
\end{eqnarray}
 which weight the QDA with different powers of the $x_i$, it is clear that
 there exists a {\em linear relation} between the {\em moments} of the QDA
 and its {\em expansion coefficients} $c_i$.
It is thus obvious that a simple matrix inversion yields the coefficients $c_i$
 as linear functions of the moments of $\phi_N$.
If the moments were known exactly for all orders, this would allow to
 reconstruct the QDA precisely.

However, the non-perturbative methods presently available for actual
 calculations, such as QCD sum rules  or QCD lattice gauge theory, can only
 provide the lowest order moments up to a quite limited accuracy of
 up to 30\%.

Furthermore, as was shown in previous works\cite{EHG94}, the relation
 is extremely
 sensitive to uncertainties or errors in the moments.
This is so because the matrix which has to be inverted, has a {\em nearly\/}
 vanishing determinant\cite{EHG94} or --~equivalently~-- some very
 small eigenvalues\cite{KiSa87}.
The problem is {\em ill-posed\/}.

The uncertainties, which can be as large as 30\%, destroy the information
 about the coefficients, e.g.\ by a sign flip.
Therefore, the model QDAs not only show very different shapes for the same
 range of the moments input, but in addition the polynomial expansion does not
 even converge.
Thus, there is an increase of the oscillations with increasing degree of the
 expansion, instead of an expected gradual refinement.

\bigskip\noindent
So what can be done in order to circumvent the above difficulties ?

\section{A simpler approach}

In order to obtain a more reasonable model of the QDA which avoids the
 problems of unphysical oscillations (due to a failure in the fine-tuning of
 higher-order expansion coefficients), we list some criteria of simplicity
 that a physical distribution amplitude should fulfill {\em in addition to
 QCD sum rules\/}\cite{EHG95}:
\begin{enumerate}
\item functional simplicity (e.g.\ an exponential ansatz)
\item minimum number of parameters
\item smooth
\item no oscillations
\item positive
\item substantially non-polynomial
\item no specific process (experiment) as input (process-independence)
\end{enumerate}
A model constructed in such a way will be called ``haplousterotic''
 (from the greek word
 $\alpha\pi\lambda o\upsilon\sigma\tau\varepsilon\rho o\varsigma$
 for ``simpler'').

\begin{figure}
\psfig{file=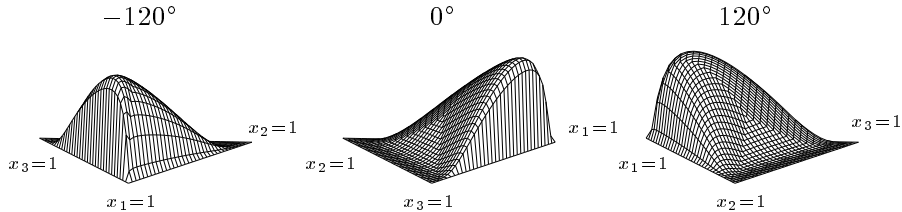,width=4.7in,silent=}
\caption{%
The haplousterotic model for the nucleon quark distribution amplitude
 shown from different directions.
This model amplitude is of exponential type --~similar to a model
 used in Ref.~\protect\citelow{BHL80}~-- and was determined using the
 QCD sum-rule moments up to third order of Ref.~\protect\citelow{COZ89}.
It reaches the edge everywhere with slope 0, although this may not be
 evident from the figure.
\label{FIGHaplousteros}}
\end{figure}

Using the above criteria, our haplousterotic model amplitude $\phi_N^{\rm Ha+}$
 was determined from the QCD sum-rule moments of COZ\cite{COZ89}.
The model is shown in Fig.~\ref{FIGHaplousteros}.
It has the form\cite{EHG95}
\begin{equation}
 \phi_N^{\rm Ha+}(x) = N \exp\left (-( {b_1^{(r)}\over x_1^r}+{b_2^{(r)}
 \over x_2^r} +{b_3^{(r)}\over x_3^r} ) \right ) \quad ,
\label{PHIHAPL}
\end{equation}
with $N=14.626$,
 $b_1^{(1)}=0.5880$, $b_2^{(1)}=8.724\times10^{-13}$ and $b_3^{(1)}=0.02413$
 for $r=1$.\footnote{There was a factor $3$ missing in the $b_i^{(r)}$
 in the first reference of Ref.~\citelow{EHG95}.}

\begin{figure}
\hbox{\hskip 6mm\psfig{file=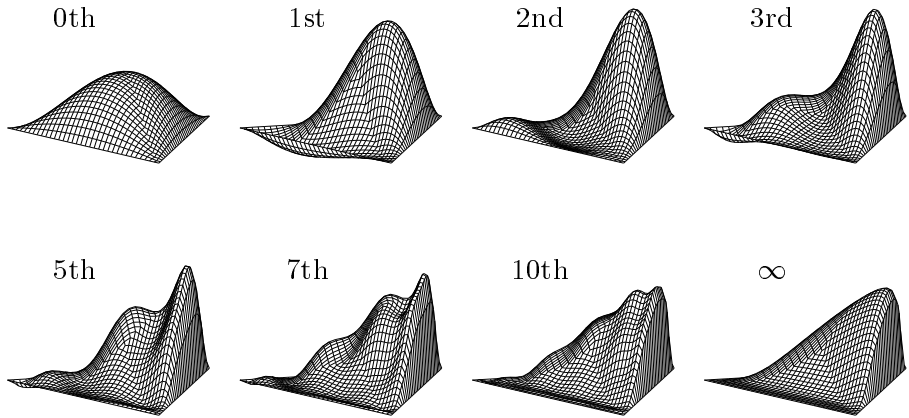,width=3.9in,silent=}}
\vglue 1.8 cm
\caption{%
The haplousterotic model amplitude is expanded into Appell polynomials.
However, to resemble the shape of the exact ($\infty$) form, polynomials
 of very high degree are needed.
\label{FIGEntwicklung}}
\end{figure}

Of course, for the purpose of investigation of the $Q^2$-evolution of the QDA
 or its convergence properties, $\phi_N^{\rm Ha+}$ can be expanded into a
 series of Appell polynomials.
A graphical representation of different orders of expansion is shown in
 Fig.~\ref{FIGEntwicklung}.
As it should be, one can observe a nice convergence.
But one also has to notice that very high polynomial degrees are needed
 for the expansion to resemble the shape of the exact model~$\phi_N^{\rm Ha+}$.
On the other hand, it is interesting to note that the second degree
 approximation to $\phi_N^{\rm Ha+}$ looks very much like the early model
 of CZ (cf.\ Fig.~\ref{FIGModelle}).

Finally, we would like to emphasize that a fixed model amplitude of the
 nucleon can be used to study systematically the effects of various 
 uncertainties and approximations 
 that are necessary to apply perturbative QCD in the calculation
 of physically relevant quantities like form factors, decay widths etc.. 
Relevant problems in this context are  e.g.\ constant or dynamical
 $\alpha_S$\cite{HEG92} and related end-point problems\cite{ILS89},
 choice of $\Lambda_{QCD}$, $k_\perp$-\cite{HEG92,BJK95} and
 Sukakov\cite{BoSt89,LiSt92}
 effects, $k_\perp$-dependence of the full wave function,
 higher-twist effects, higher-order~$\alpha_S$ etc.

\section{Summary}

Due to the sensitivity of the relation between moments and coefficients to
 small variations in the moments (which are unavoidable when the moments
 are calculated, e.g.\ by the method of QCD sum rules) strong oscillations
 result in the polynomial approximation, which lead to a non-convergence. 

A way out of this dilemma is to restrict the structure of the quark
 distribution amplitude to physical and phenomenological criteria of simplicity
 unless significantly more accurate methods for obtaining QCD wave functions
 are available. 

Our haplousterotic model of the nucleon QDA contains all available and
 physically relevant information and it will be an interesting challenge
 to test it in VCS.

\section*{Acknowledgments}
Two of us (RE and JH) would like to thank the organizers of the workshop on
 ``Virtual Compton Scattering 96'' in Clermont-Ferrand for their kind
 hospitality during the workshop.
We are also grateful for critical remarks and comments to V.M. Braun,
 V.L. Chernyak and A.V. Radyushkin.

\end{document}